Relativistic four-quark equations and cryptoexotic mesons spectroscopy.

Gerasyuta S.M.*, Kochkin V.I.
Department of Theoretical Physics, St. Petersburg State University, 198904, St. Petersburg, Russia.

Summary. The four-quark equations are found in the framework of the dispersion relation technique. The approximate solutions of these equations using the method based on the extraction of leading singularities of the amplitudes are obtained. The four-quark amplitudes of cryptoexotic mesons including the quarks of three flavours (u, d, s) are calculated. The mass values of low-lying cryptoexotic mesons are calculated.

---

*   Present address: Department of Physics, LTA, Institutski
Per. 5, St. Petersburg 194021, Russia

## 1. Introduction.

The history of $qq\bar{q}\bar{q}$ states dates back to 1967 when Astier [1] suggested that $K\bar{K}$ bound states could explain the low mass $I = 1$ enhancements seen in $p\bar{p} \to K\bar{K}\pi$. In 1968, Rosner [2] and Harari [3] noticed that the dual t-channel of baryon-antibaryon scattering through meson exchanges represents $qq\bar{q}\bar{q}$ system, "baryonia". The first extensive calculation was by Jaffe [4] who applied a semi-classical approximation of the MIT-bag model to $qq\bar{q}\bar{q}$ spectroscopy. He noted that colour magnetic interactions drive the multiplet splitting and he suggested that the lowest scalar nonet were $qq\bar{q}\bar{q}$ bound states, hence explaining many of their peculiar properties.

In 1977, Chan and Hogaasen [5] used bag and string models to investigate baryonia in their unmixed colour states. They predicted a tower of $(qq)_6 \otimes (\bar{q}\bar{q})_{\bar{6}}$ S-wave states held apart by a high angular momentum barrier. In 1980 Wicklund [6] suggested that an isoscalar $K\bar{K}$ bound state explained $\pi\pi \to f_0(975) \to K\bar{K}$ data. Weinstein and Isgur [7] have performed extensive calculations in the four-quark system with the quark potential model. Their results suggest that the $a_0(980)$ and $f_0(975)$ are indeed $K\bar{K}$ bound states but that no other $qq\bar{q}\bar{q}$ states exist (at least in the sectors they studied). These authors noted the mounting evidence in favour of the molecular interpretation of the $a_0$ and $f_0$. For example, the $qq\bar{q}\bar{q}$ bound state scenario provides an explanation for the narrow width of these particles. Barnes [8] has argued that the $\gamma\gamma$-decay widths imply that the $a_0$ and $f_0$ are $qq\bar{q}\bar{q}$ states rather than $q\bar{q}$ states. Finally, the near degenerate of the $a_0$ and $f_0$ suggest that if they are $q\bar{q}$ states then they are approximately ideally mixed and hence the ratio of decay width to $\pi\pi$ should be roughly 4 rather than the observed value of 1/2 [9].

The bag model [10, 11] provides another important vehicle for investigating multiquark physics. It has the benefit of utilizing an attractive physical picture of confinement, it incorporates asymptotic freedom by describing the quarks inside of the bag with the free Dirac equation.

Jaffe and Low introduced a quantity they called the P-matrix whose poles are expected to correspond to the energies calculated in the bag model [12]. Unfortunately their program is computationally difficult.

It should be stressed that all predictions of multiquark bound states are entirely model-dependent, thus the predictions of the quark potential model serve as an important test [13, 14].

In recent papers [15, 16] the relativistic Faddeev equations are represented in the form of the dispersion over the two-body subenergy. The behavior of the low-energy three-particle amplitude is determined by its leading singularities in the pair invariant masses. The suggested method of approximate solution of the relativistic Faddeev equation was verified on the example of the S-wave baryons. The calculated mass values of two baryonic multiplets $J^P = \frac{1}{2}^+, \frac{3}{2}^+$ are in good agreement with the experimental ones. The dispersion relations technique allows us to consider relativistic effects in the composite systems. Double dispersion relations over the masses of the composite particles were used in [16] for the consideration of the form factors. The behavior of the nucleon electromagnetic form factors at small and intermediate momentum transfer $Q^2 < 1,5 \, GeV^2$ are determined.

In the present paper the relativistic four-quark equations are constructed in the form of the dispersion relation over the two-body subenergy. We calculated the masses of the cryptoexotic multiplets using the method based on the extraction of leading singularities of the amplitude.

In Section 2 the relativistic four-quark equations are constructed in the form of the dispersion relation over the two-body subenergy. The approximate solutions of these equations using the method based on the extraction of leading singularities of the amplitude are obtained. The four-quark amplitudes of cryptoexotic mesons are calculated.

Section 3 is devoted to the calculation results for the cryptoexotic meson mass spectrum (Table1).

Table 2 shows the contributions of different subamplitudes to the four-quark amplitude.

In the Conclusion the status of the considered model is discussed.

In the Appendix A the quark-antiquark vertex functions and the phase spaces for the cryptoexotic mesons are given (Tables 3, 4).

In the Appendix B we search the integration contours of functions $J_1$, $J_2$, $J_3$, which are determined by the interaction of the four quarks.

## 2. Four-quark amplitudes of the cryptoexotic mesons.

In the recent papers [17, 18] the relativistic four-quark equations are constructed. We consider only the planar diagrams, the other diagrams due to the rules of $1/N_c$ - expansion [19, 20] are neglected. For the sake of simplicity one considered the case of the $SU(3)_f$ -symmetry that the masses of all particles are equal.

In the present paper we investigate scattering amplitudes of the constituent quarks of three flavours (u, d, s). The poles of these amplitudes determine the masses of the cryptoexotic mesons. The masses of the constituent quarks $u$ and $d$ are of the order of 300-400 MeV, the strange quark is 100-150 MeV heavier. The strange quark violates the flavour $SU(3)_f$ -symmetry. The constituent quark is color triplet and quark amplitudes obey the global color symmetry.

We derived the relativistic four-quark equations in the framework of the dispersion relation technique. Let some current produce two pairs of quark-antiquark (Fig.1,2). The consideration of diagrams in Fig.1,2 allows to present graphically the equations for the four-quark amplitudes. However, the correct equations for the amplitude are obtained at the account of all possible subamplitudes. It corresponds to the division complete system on the subsystems from the smaller number of particles. Then one should present four-particle amplitude as a sum of six subamplitudes: $A = A_{12} + A_{13} + A_{14} + A_{23} + A_{24} + A_{34}$. It defines the division of the diagrams in the groups according to the last interaction of particles. In this case we need to consider only one group of diagrams and the amplitude corresponding to them, for example $A_{12}$. We must take into account each sequence of the inclusion of interaction. For instance, the process beginning with interaction of the particles 1 and 2 can proceed by the three ways: particle 3 and 4 consistently join a chosen pair, or begin to interact among themselves, and each of the three ways of the connection there should correspond to their own amplitudes [21, 22]. Therefore the diagrams corresponding to amplitude $A_{12}$ are divided in three group $A_1(s, s_{12}, s_{123})$, $A_1(s, s_{12}, s_{124})$ and $A_2(s, s_{12}, s_{34})$ (moreover the subamplitudes $A_1(s, s_{12}, s_{123})$ and $A_1(s, s_{12}, s_{124})$ are analogous).

In the present paper one supposed that the transitions $q_a \bar{q}_a \to q_b \bar{q}_b$ (a, b = u, d, s) for the colour singlet states are absent [4]. The similar considerations are not valid for the colour octet states.

The equations for the four-quark amplitudes in the graphic form are presented (Fig.1,2). The coefficients are determined by the permutation of quarks [21]. The four-quark amplitude $A$ is defined by contribution of equal subamplitudes $A_{12} = A_{34}$. The other subamplitudes $A_{13}$, $A_{24}$ (describing the interactions of diquarks) and $A_{14}$, $A_{23}$ (satisfying the Okubo - Zweig - Iisuka rule) do not contribute to the amplitude $A$. Then we can consider only one group of diagrams corresponding to the amplitude $A_{12}$.

In order to present the amplitudes $A_l(s,s_{12},s_{123})$ ($l$=1-4) and $A_p(s,s_{12},s_{34})$ (p=5-7) in form of the dispersion relation it is necessary to define the amplitude of two-quark interaction $a_j(s_{ik})$. One uses the results of the bootstrap quark model [23,24] and writes down the pair quarks amplitude in the form:

(1) $\quad a_j(s_{ik}) = \dfrac{G_j^2(s_{ik})}{1 - B_j(s_{ik})},$

(2) $\quad B_j(s_{ik}) = \displaystyle\int_{(m_i+m_k)^2}^{\Lambda_{ik}} \dfrac{ds'_{ik}}{\pi} \dfrac{\rho_j(s'_{ik}) G_j^2(s'_{ik})}{s'_{ik} - s_{ik}},$

Here $G_j(s_{ik})$ is the quark-antiquark vertex function. $B_j(s_{ik})$, $\rho_j(s_{ik})$ are the Chew-Mandelstam function [25] and the phase space respectively. We introduced the cut-off parameter $\Lambda_{ik}$. There j=1 corresponds to pair of quarks $q\bar{q}$ with $J^{PC}= 0^{++}$, $1^{++}$, $2^{++}$, $0^{-+}$, $1^{--}$, $1^{+-}$ (colour singlet $SU(3)_c$) and j=2 defines the quark pair with $J^{PC} = 1^{--}$ in colour channel $8_c$ (constituent gluon). The vertex functions are shown in the Table 3, the functions $\rho_j(s_{ik})$ are given in the Appendix A (see Table 4). In the case in question the interacting quarks do not produce bound state, then the integration in (3)-(9) is carried out from the threshold $(m_i + m_k)^2$ to the cut-off $\Lambda_{ik}$. The integral equation systems, corresponding to Fig 1,2, have the following form:

(3) $\quad A_1(s,s_{12},s_{123}) = \dfrac{\lambda_1 B_1^{uu}(s_{12})}{1 - B_1^{uu}(s_{12})} + 2\dfrac{G_1(s_{12})}{1 - B_1^{uu}(s_{12})}[\hat{J}_1 A_1(s,s'_{13},s_{123}) + \hat{J}_3 A_6(s,s'_{13},s'_{24})]$

(4) $\quad A_2(s,s_{12},s_{123}) = \dfrac{\lambda_2 B_1^{ss}(s_{12})}{1 - B_1^{ss}(s_{12})} + 2\dfrac{G_1(s_{12})}{1 - B_1^{ss}(s_{12})}[\hat{J}_1^{ss} A_2(s,s'_{13},s_{123}) + \hat{J}_3^{ss} A_5(s,s'_{13},s'_{24})]$

(5) $\quad A_3(s,s_{12},s_{123}) = \dfrac{\lambda_3 B_1^{us}(s_{12})}{1 - B_1^{us}(s_{12})} + 2\dfrac{G_1(s_{12})}{1 - B_1^{us}(s_{12})} \hat{J}_3^s A_7(s,s'_{13},s'_{24})$

(6) $\quad A_4(s,s_{12},s_{123}) = \dfrac{\lambda_4 B_1^{ud}(s_{12})}{1 - B_1^{ud}(s_{12})} + 2\dfrac{G_1(s_{12})}{1 - B_1^{ud}(s_{12})} \hat{J}_3 A_6(s,s'_{13},s'_{24})$

(7) $\quad A_5(s,s_{12},s_{34}) = \dfrac{\lambda_5 B_2^{ss}(s_{12}) B_2^{ss}(s_{34})}{[1 - B_2^{ss}(s_{12})][1 - B_2^{ss}(s_{34})]} + 4\dfrac{G_2(s_{12}) G_2(s_{34})}{[1 - B_2^{ss}(s_{12})][1 - B_2^{ss}(s_{34})]} \times$
$\times [\hat{J}_2 A_1(s,s'_{13},s'_{134}) + \hat{J}_2 A_4(s,s'_{13},s'_{134}) + \hat{J}_2^s A_3(s,s'_{13},s'_{134}) + \hat{J}_2^{ss} A_2(s,s'_{13},s'_{134})]$

(8) $\quad A_6(s,s_{12},s_{34}) = \dfrac{\lambda_6 B_2^{uu}(s_{12}) B_2^{uu}(s_{34})}{[1 - B_2^{uu}(s_{12})][1 - B_2^{uu}(s_{34})]} + 4\dfrac{G_2(s_{12}) G_2(s_{34})}{[1 - B_2^{uu}(s_{12})][1 - B_2^{uu}(s_{34})]} \times$
$\times [\hat{J}_2 A_1(s,s'_{13},s'_{134}) + \hat{J}_2 A_4(s,s'_{13},s'_{134}) + \hat{J}_2^s A_3(s,s'_{13},s'_{134}) + \hat{J}_2^{ss} A_2(s,s'_{13},s'_{134})]$

(9) $\quad A_7(s,s_{12},s_{34}) = \dfrac{\lambda_7 B_2^{ss}(s_{12}) B_2^{uu}(s_{34})}{[1 - B_2^{ss}(s_{12})][1 - B_2^{uu}(s_{34})]} + 4\dfrac{G_2(s_{12}) G_2(s_{34})}{[1 - B_2^{ss}(s_{12})][1 - B_2^{uu}(s_{34})]} \times$
$\times [\hat{J}_2 A_1(s,s'_{13},s'_{134}) + \hat{J}_2 A_4(s,s'_{13},s'_{134}) + \hat{J}_2^s A_3(s,s'_{13},s'_{134}) + \hat{J}_2^{ss} A_2(s,s'_{13},s'_{134})] ,$

$\lambda_i$ are the current constants. Here we introduce the integral operators:

(10) $\quad \hat{J}_1(s;m_1;m_2;m_3;m_4) = \displaystyle\int_{(m_1+m_2)^2}^{\Lambda_{12}} \dfrac{ds'_{12}}{\pi} \dfrac{\rho_1(s'_{12}) \cdot G_1(s'_{12})}{s'_{12} - s_{12}} \int_{-1}^{+1} \dfrac{dz_1}{2}$

(11) $\quad \hat{J}_2(s;m_1;m_2;m_3;m_4) = \displaystyle\int_{(m_1+m_2)^2}^{\Lambda_{12}} \dfrac{ds'_{12}}{\pi} \dfrac{\rho_2(s'_{12}) \cdot G_2(s'_{12})}{s'_{12} - s_{12}} \int_{(m_3+m_4)^2}^{\Lambda_{34}} \dfrac{ds'_{34}}{\pi} \dfrac{\rho_2(s'_{34}) \cdot G_2(s'_{34})}{s'_{34} - s_{34}} \int_{-1}^{+1} \dfrac{dz_3}{2} \int_{-1}^{+1} \dfrac{dz_4}{2}$

$$\hat{J}_3(s; m_1; m_2; m_3; m_4) = \frac{1}{4\pi} \times$$

(12)
$$\times \int_{(m_1+m_2)^2}^{\Lambda_{12}} \frac{ds'_{12}}{\pi} \frac{\rho_1(s'_{12}) \cdot G_1(s'_{12})}{s'_{12} - s_{12}} \int_{-1}^{+1} \frac{dz_1}{2} \int_{-1}^{+1} dz \int_{z_2^-}^{z_2^+} dz_2 \frac{1}{\sqrt{1 - z^2 - z_1^2 - z_2^2 + 2zz_1z_2}},$$

One used the following indications:

$\hat{J}_i(s; m; m; m; m) = \hat{J}_i$

$\hat{J}_i(s; m_s; m_s; m_s; m_s) = \hat{J}_i^{ss}$, here i=1,2,3

$\hat{J}_2(s; m; m; m_s; m_s) = \hat{J}_2^s$, $\hat{J}_3(s; m; m_s; m; m_s) = \hat{J}_3^s$,

there $m$ and $m_s$ are the masses of nonstrange and strange quarks respectively. The $B^{qq}$-functions are defined by the composition of $q\bar{q}$-quark pair for the final state. In the equations (10) and (12) $z_1$ is the cosine of the angle between the relative momentum of the particles 1 and 2 in the intermediate state and that of the particle 3 in the final state, which is taken in the c.m. of particles 1 and 2. In the equation (12) $z$ is the cosine of the angle between the momentum of the particles 3 and 4 in the final state, which is taken in the c.m. of particles 1 and 2. $z_2$ is the cosine of the angle between the relative momentum of particles 1 and 2 in the intermediate state and the momentum of the particle 4 in the final state, which is taken in the c.m. of particles 1 and 2. In the equation (11) we have defined: $z_3$ is the cosine of the angle between relative momentum of particles 1 and 2 in the intermediate state and that of the relative momentum of particles 3 and 4 in the intermediate state, which is taken in the c.m. of particles 1 and 2; $z_4$ is the cosine of the angle between the relative momentum of the particles 3 and 4 in the intermediate state and that of the momentum of the particle 1 in the intermediate state which is taken in the c.m. of particles 3, 4. Using (13)-(17) we can pass from the integration over the cosines of the angles to the integration over the subenergies. The choice of integration contours of functions $J_1$, $J_2$, $J_3$ do not differ from the papers [17, 18] (see Appendix B).

(13)
$$s'_{13} = m_1^2 + m_3^2 + \frac{(s_{123} - s'_{12} - m_3^2)(s'_{12} - m_2^2 + m_1^2)}{2s'_{12}} +$$
$$+ \frac{z_1}{2s'_{12}} \sqrt{[(s_{123} - s'_{12} - m_3^2)^2 - 4s'_{12}m_3^2][(s'_{12} - m_2^2 + m_1^2)^2 - 4s'_{12}m_1^2]}$$

(14)
$$s'_{24} = m_2^2 + m_4^2 + \frac{(s'_{124} - s'_{12} - m_4^2)(s'_{12} - m_1^2 + m_2^2)}{2s'_{12}} +$$
$$+ \frac{z_2}{2s'_{12}} \sqrt{[(s'_{124} - s'_{12} - m_4^2)^2 - 4s'_{12}m_4^2][(s'_{12} - m_1^2 + m_2^2)^2 - 4s'_{12}m_2^2]}$$

(15) $$z = \frac{2s'_{12}(s + s'_{12} - s_{123} - s'_{124}) - (s_{123} - s'_{12} - m_3^2)(s'_{124} - s'_{12} - m_4^2)}{\sqrt{[(s_{123} - s'_{12} - m_3^2)^2 - 4m_3^2 s'_{12}][(s'_{124} - s'_{12} - m_4^2)^2 - 4m_4^2 s'_{12}]}}$$

(16) $$s'_{134} = m_1^2 + s'_{34} + \frac{s - s'_{12} - s'_{34}}{2} + \frac{z_3}{2} \sqrt{\frac{s'_{12} - 4m_1^2}{s'_{12}} [(s - s'_{12} - s'_{34})^2 - 4s'_{12}s'_{34}]}$$

(17) $$s'_{13} = m_1^2 + m_3^2 + \frac{s'_{134} - s'_{34} - m_1^2}{2} + \frac{z_4}{2} \sqrt{\frac{s'_{34} - 4m_3^2}{s'_{34}} [(s'_{134} - s'_{34} - m_1^2)^2 - 4m_1^2 s'_{34}]}$$

Let us extract two-particle singularities in the amplitudes $A_l(s, s_{12}, s_{123})$ and $A_p(s, s_{12}, s_{34})$:

(18) $A_l(s, s_{12}, s_{123}) = \dfrac{\alpha_l(s, s_{12}, s_{123}) B_1(s_{12})}{1 - B_1(s_{12})}$

(19) $A_p(s, s_{12}, s_{34}) = \dfrac{\alpha_p(s, s_{12}, s_{34}) B_2(s_{12}) B_2(s_{34})}{[1 - B_2(s_{12})][1 - B_2(s_{34})]}$,

$l = 1\text{-}4$, p=5-7.

In the amplitude $A_l(s, s_{12}, s_{123})$ we do not extract three-particle singularity, because it is weaker than two-particle and taking into account in the function $\alpha_l(s, s_{12}, s_{123})$.

We used the classification of singularities, which was proposed in papers [17, 18]. The construction of approximate solution of the (18) and (19) is based on the extraction of the leading singularities of the amplitudes. The main singularities in $s_{ik}$ are from pair rescattering of the particles i and k. First of all there are threshold square root singularities. Also possible are pole singularities which correspond to the bound states. They are situated on the first sheet of complex $s_{ik}$ plane in case of real bound state and on the second sheet in case of virtual bound state. The diagrams Fig.1, 2 apart from two-particle singularities have their specific triangular singularities and the singularities correspond to the interaction of four particles. Such classification allows us to search the approximate solution of (18) and (19) by taking into account some definite number of leading singularities and neglecting all the weaker ones. We consider the approximation, which corresponds to the single interaction of all four particles (two-particle, triangle and four-particle singularities). The functions $\alpha_l(s, s_{12}, s_{123})$ and $\alpha_p(s, s_{12}, s_{34})$ are the smooth functions of $s_{ik}$, $s_{ijk}$ as compared with the singular part of the amplitudes, hence they can be expanded in a series in the singularity point and only the first term of this series should be employed further. Using this classification one define the functions $\alpha_l(s, s_{12}, s_{123})$ and $\alpha_p(s, s_{12}, s_{34})$ as well as the B-functions in the middle point of the physical region of Dalitz-plot at the point $s^0$:

(20) $\begin{aligned}
s_{1i}^0 &= s_{j4}^0 = \dfrac{s + 8m^2}{6} \\
s^0 &= \dfrac{4[s - 2s_{1i}^0 + 2(m_1^2 + m_2^2 + m_3^2 + m_4^2)]}{(m_1 + m_j)^2 + (m_1 + m_4)^2 + (m_2 + m_3)^2 + (m_i + m_4)^2} \\
s_{123} &= s_{1i}^0 + \tfrac{1}{4}(m_1 + m_j)^2 s^0 + \tfrac{1}{4}(m_2 + m_3)^2 s^0 - m_1^2 - m_2^2 - m_3^2
\end{aligned}$

Here one suggested i=2, j=3 for $\hat{J}_1$ and $\hat{J}_2$; i=3, j=2 for $\hat{J}_3$. Moreover, the other choice of the middle point does not change essentially the obtained results. Such a choice of points $s_{12}^0$, $s_{34}^0$ allows as to replace the integral equations (3)-(9) by the algebraic equations (21)-(27) respectively:

(21) $\alpha_1 = \lambda_1 + 2[\alpha_1 J_1 + \alpha_6 J_3] / B_1^{uu}(s_{12}^0)$

(22) $\alpha_2 = \lambda_2 + 2[\alpha_2 J_1^{ss} + \alpha_5 J_3^{ss}] / B_1^{ss}(s_{12}^0)$

(23) $\alpha_3 = \lambda_3 + 2\alpha_7 J_3^s / B_1^{us}(s_{12}^0)$

(24) $\alpha_4 = \lambda_4 + 2\alpha_6 J_3 / B_1^{ud}(s_{12}^0)$

(25) $\alpha_5 = \lambda_5 + 4[(\alpha_1 + \alpha_4) J_2 + \alpha_3 J_2^s + \alpha_2 J_2^{ss}] / [B_2^{ss}(s_{12}^0) B_2^{ss}(s_{34}^0)]$

(26) $\alpha_6 = \lambda_6 + 4[(\alpha_1 + \alpha_4) J_2 + \alpha_3 J_2^s + \alpha_2 J_2^{ss}] / [B_2^{uu}(s_{12}^0) B_2^{uu}(s_{34}^0)]$

(27) $\alpha_7 = \lambda_7 + 4[(\alpha_1 + \alpha_4) J_2 + \alpha_3 J_2^s + \alpha_2 J_2^{ss}] / [B_2^{us}(s_{12}^0) B_2^{us}(s_{34}^0)]$

Here we introduce following functions:

$$(28) \quad J_1(s; m_1; m_2; m_3; m_4) = G_1^2 B_1(s_{13}^0) \int_{(m_1+m_2)^2}^{\Lambda_{12}} \frac{ds'_{12}}{\pi} \frac{\rho_1(s'_{12})}{s'_{12} - s^0_{12}} \int_{-1}^{+1} \frac{dz_1}{2} \frac{1}{1 - B_1(s'_{13})}$$

$$(29) \quad J_2(s; m_1; m_2; m_3; m_4) = G_2^4 B_1(s_{13}^0) \times$$
$$\times \int_{(m_1+m_2)^2}^{\Lambda_{12}} \frac{ds'_{12}}{\pi} \frac{\rho_2(s'_{12})}{s'_{12} - s^0_{12}} \int_{(m_3+m_4)^2}^{\Lambda_{34}} \frac{ds'_{34}}{\pi} \frac{\rho_2(s'_{34})}{s'_{34} - s^0_{34}} \int_{-1}^{+1} \frac{dz_3}{2} \int_{-1}^{+1} \frac{dz_4}{2} \frac{1}{1 - B_1(s'_{13})}$$

$$(30) \quad J_3(s; m_1; m_2; m_3; m_4) = G_1^2 B_2(s_{13}^0) B_2(s_{24}^0) \frac{1 - B_1(s_{12}^0, \Lambda_{12})}{1 - B_1(s_{12}^0, \widetilde{\Lambda}_{12})} \frac{1}{4\pi} \times$$
$$\times \int_{(m_1+m_2)^2}^{\widetilde{\Lambda}_{12}} \frac{ds'_{12}}{\pi} \frac{\rho_1(s'_{12})}{s'_{12} - s^0_{12}} \int_{-1}^{+1} \frac{dz_1}{2} \int_{-1}^{+1} dz \int_{z_2^-}^{z_2^+} dz_2 \frac{1}{\sqrt{1 - z^2 - z_1^2 - z_2^2 + 2z z_1 z_2}} \frac{1}{[1 - B_2(s'_{13})][1 - B_2(s'_{24})]}$$

As the integration region the physical region of the reaction should be chosen, therefore $-1 \leq z_i \leq 1$ ( i=1,2,3,4 ). From these conditions we can define the regions of the integration over $s'_{13}$, $s'_{24}$, $s'_{134}$, $s'_{124}$. Let us consider the integration region over $s'_{124}$. For this purpose we use equation (15). This condition corresponds to $0 \leq z^2 \leq 1$. By consideration of this inequality one can obtain:

$$(31) \quad s'^{\pm}_{124} = s'_{12} + m_4^2 + \frac{(s - s_{123} - m_4^2)(s_{123} + s'_{12} - m_3^2)}{2 s_{123}} \pm$$
$$\pm \frac{1}{2 s_{123}} \sqrt{[(s_{123} - s'_{12} - m_3^2)^2 - 4 m_3^2 s'_{12}][(s - s_{123} - m_4^2)^2 - 4 m_4^2 s_{123}]}$$

We must take into account the upper restriction of the integration region over $s'_{12}$ in $J_3$:

$$(32) \quad \widetilde{\Lambda}_{12} = \begin{cases} \Lambda_{12}, & if \ \Lambda_{12} \leq (\sqrt{s_{123}} + m_3)^2 \\ (\sqrt{s_{123}} + m_3)^2, & if \ \Lambda_{12} > (\sqrt{s_{123}} + m_3)^2 \end{cases}$$

The integration contours of the functions $J_1$, $J_2$, $J_3$ are given in the Appendix B. The function $J_3$ takes into account the singularity, which corresponds to the simultaneous vanishing of all propagators in the four-particle diagram like those in Fig.1. In the case in question the functions $\alpha_i(s)$ are determined as:

(33) $\quad \alpha_i(s) = F_i(s, \lambda_i) / \Delta(s)$

There $\Delta(s)$ is the determinant:

(34) $\quad \Delta(s) = (1 - 8 J_2 J_3 - 8 J_2^s J_3^s)(1 - 2 J_1)(1 - 2 J_1^{ss}) - 8 J_2 J_3 (1 - 2 J_1^{ss}) - 8 J_2^{ss} J_3^{ss} (1 - 2 J_1)$

The right-hand sides of (33) might have a pole in $s$ which corresponds to the bound state of the four quarks. The poles of rescattering amplitudes for the cryptoexotic mesons $J^{PC} = 0^{++}$, $1^{++}$, $2^{++}$, $0^{-+}$, $1^{--}$, $1^{+-}$ correspond to the bound state and determine the masses of the cryptoexotic mesons.

## 3. Calculation results.

In the bootstrap quark of model [23, 24] there is a bound state in the gluon channel with mass of the order of 0.7 GeV. This bound state should be identified as a constituent gluon. It should be pointed out that such a value of the mass agrees with the hard-process phenomenology [26]. An analogous estimation of the gluon mass is obtained in the bag models [27, 28].

The strange quark in our model gives rise to the violation of the flavour $SU(3)_f$-symmetry. In order to avoid an additional violation parameter we introduce the scale shift of the dimensional parameter: $\Lambda_{ik} = (m_i + m_k)^2 \Lambda / 4$.

In the considered calculation the quark masses ($m$ and $m_s$) are not fixed. In order to fix anyhow $m$ and $m_s$ we assume $m = 368\, MeV$ ($m \geq \frac{1}{4} m_{f_2}(1430)$) and $m_s = 513\, MeV$ ($m_s \geq \frac{1}{4} m_{f_2}(2010)$).

The model under consideration proceeds from the assumption that the quark interaction forces are the two-component ones. The long-range component is due to the confinement. When the low-lying mesons are considered, the long-range component of the forces is neglected. The creation of ordinary mesons is mainly due to the constituent gluon exchange (Fig. 3a). But for the excited mesons the long-range forces are important. Namely, the confinement of the $q\bar{q}$ pair with comparatively large energy is actually realized as the production of the new $q\bar{q}$ pairs. This means that in the transition $q\bar{q} \to q\bar{q}$ the forces appear which are connected with the process of the Fig. 3b type. These box-diagrams can be important in the formation of hadron spectra [29]. We do not see any difficult in taking into account the box-diagrams with the help of the dispersion technique. For the sake of simplicity we restrict ourselves to the introduction of quark mass shift $\Delta$, which are defined by the contributions of the nearest production thresholds of pair mesons $\pi\pi$, $\pi\eta$, $K\bar{K}$, $K\eta$ and so on. We suggest that the parameter $\Delta$ takes into account the confinement potential effectively: $m^{eff} = m + \Delta$, $m_s^{eff} = m_s + \Delta$ and changes the behavior of pair quarks amplitude (1). It allows us to construct the excited cryptoexotic mesons amplitudes and calculate the mass spectrum by analogy with the P-wave meson spectrum in the bootstrap quark model [30]. The model in consideration have another two parameters: cut-off parameter $\Lambda$ and gluon constant $g$ for each group of mesons. The subenergy cut-off $\Lambda$ and the vertex function $g$ can be determined by mean of fixing of cryptoexotic mesons mass values ($J^{PC} = 0^{-+}, 2^{++}$). The vertex functions of various types of the interactions are given in Table 3. The calculated values of mass cryptoexotic mesons (groups I-III) are shown in the Table 1. The results are in good agreement with the experimental data [31] and the other model results [32-37]. However, for the lowest $J^{PC} = 0^{++}$ multiplet the discrepancy between calculated and observed values of masses $M_{f_0} = 1010\,(985)\,\text{MeV}$ is more than others. It is possible that this is due to the admixture of the $q\bar{q}$ scalar state, moreover the mass value of two-quark state with $J^{PC} = 0^{++}$ in the bootstrap quark model $M_{f_0} = 870\,(985)\,\text{MeV}$ is obtained.

## 4. Conclusion.

In the present paper in the framework of approximate method of solution four-particle relativistic problem the mass spectrum of cryptoexotic mesons, including u, d, s - quarks, are calculated. The mass values of already detected candidates of cryptoexotic mesons are calculated. The interesting result of this model is the calculation of cryptoexotic meson amplitudes, which contain the contributions of seven subamplitudes: four four-quark amplitudes and three glueball amplitudes. The contributions of glueball subamplitudes (groups I-III) are given in Table 2. The decay width of cryptoexotic mesons can be calculated in the framework this model. The suggested approximate method allows to construct the four-quark amplitudes, including heavy quarks Q=c, b and calculate the mass spectrum of heavy cryptoexotic mesons.

****

The authors would like to thank A.A. Andrianov, V.A. Franke and Yu.V. Novozhilov for useful discussions.

## APPENDIX A

The two-particle phase space for the unequal quark masses is defined as:

$$\rho_1(s_{ik}, J^{PC}) = \left( \alpha(J^{PC}) \frac{s_{ik}}{(m_i + m_k)^2} + \beta(J^{PC}) + \delta(J^{PC}) \frac{(m_i - m_k)^2}{s_{ik}} \right) \times$$

$$\times \frac{\sqrt{[s_{ik} - (m_i + m_k)^2][s_{ik} - (m_i - m_k)^2]}}{s_{ik}}$$

$$\rho_2(s_{ik}) = \rho_1(s_{ik}, 1^{--})$$

$$B_2(s_{ik}) = \frac{1}{3}[2B_2(s_{ik}, m, m) + B_2(s_{ik}, m_s, m_s)]$$

The coefficients $\alpha(J^{PC})$, $\beta(J^{PC})$ and $\delta(J^{PC})$ are given in Table 4.

## APPENDIX B

The integration contour 1 (Fig.4) corresponds to the connection $s_{123} < (\sqrt{s_{12}} - m_3)^2$, the contour 2 is defined by the connection $(\sqrt{s_{12}} - m_3)^2 < s_{123} < (\sqrt{s_{12}} + m_3)^2$. The point $s_{123} = (\sqrt{s_{12}} - m_3)^2$ is not singular, that the round of this point at $s_{123} + i\varepsilon$ and $s_{123} - i\varepsilon$ gives identical result. $s_{123} = (\sqrt{s_{12}} + m_3)^2$ is the singular point, but in our case the integration contour can not pass through this point that the region in consideration is situated below the production threshold of the four particles $s < 16m^2$. The similar situation for the integration over $s_{13}$ in the function $J_3$ is occurred. But the difference consists of the given integration region that is conducted between the complex conjugate points (contour 2 Fig.4). In Fig.4, 5b, 6 the dotted lines define the square root cut of the Chew-Mandelstam functions. They correspond to two-particles threshold and also three-particles threshold in Fig.5a. The integration contour 1 (Fig.5a) is determined by $s < (\sqrt{s_{12}} - \sqrt{s_{34}})^2$, the contour 2 corresponds to the case $(\sqrt{s_{12}} - \sqrt{s_{34}})^2 < s < (\sqrt{s_{12}} + \sqrt{s_{34}})^2$. $s = (\sqrt{s_{12}} - \sqrt{s_{34}})^2$ is not singular point, that the round of this point at $s+i\varepsilon$ and $s-i\varepsilon$ gives identical results. The integration contour 1 (Fig.5b) is determined by region $s<(\sqrt{s_{12}}-\sqrt{s_{34}})^2$ and $s_{134}<(\sqrt{s_{34}}-m_1)^2$, the integration contour 2 corresponds to $s<(\sqrt{s_{12}}-\sqrt{s_{34}})^2$ and $(\sqrt{s_{34}}-m_1)^2 \leq s_{134}<(\sqrt{s_{34}}+m_1)^2$. The contour 3 is defined by $(\sqrt{s_{12}}-\sqrt{s_{34}})^2 < s < (\sqrt{s_{12}}+\sqrt{s_{34}})^2$. Here the singular point would be $s_{134}=(\sqrt{s_{34}}+m_1)^2$. But in our case this point is not achievable, if one has the condition $s<16m^2$. We have to consider the integration over $s_{24}$ in the function $J_3$. While $s_{124} < s_{12}+m_4^2+4m_2m_4s_{12}/(s_{12}-m_1^2+m_2^2)$ the integration is conducted along the complex axis (the contour 1, Fig.6). If we come to the point $s_{124}=s_{12}+m_4^2+4m_2m_4s_{12}/(s_{12}-m_1^2+m_2^2)$, that the output into the square root cut of Chew-Mandelstam function (contour 2, Fig.6) is occurred. In this case the part of the integration contour in nonphysical region is situated and the integration contour along the real axis is conducted. The other part of integration contour corresponds to physical regions. This part of integration contour along the complex axis is conducted. The suggested calculation show that the contribution of the integration over the nonphysical region is small.

Table 1. Cryptoexotic low-lying meson masses.

| $J^{PC}$ | masses MeV | | |
|---|---|---|---|
| | I | II | III |
| $0^{++}$ | 1010 $f_0$ (985) | 1418 $f_0$ (1370) | 1575 $f_0$ (1500) |
| $1^{++}$ | 1306 $f_1$ (1285) | 1443 $f_1$ (1420) | 1615 $f_1$ (1530) |
| $2^{++}$ | 1430 $f_2$ (1430) | 1520 $f_2$ (1520) | 1710 $f_2$ (1710) |
| $0^{-+}$ | 958 $\eta$ (958) | 1295 $\eta$ (1295) | 1440 $\eta$ (1440) |
| $1^{--}$ | 1416 $\rho$ (1450) | 1511 ( - ) | 1697 ( - ) |
| $1^{+-}$ | 958 ( - ) | 1295 ( - ) | 1440 ( - ) |

Parameters of model: I) $\Lambda$ =32,7; $g$ =0,1613; $\Delta$ =0, II) $\Lambda$ =18; $g$ =0,2535; $\Delta$ =22,5 MeV, III) $\Lambda$ =19,6; $g$ =0,2354; $\Delta$ =70 MeV. Experimental values of the cryptoexotic mesons are given in parentheses [31].

Table 2. Contributions of glueball $A_5+A_6+A_7$ subamplitudes to the cryptoexotic meson amplitude in % (groups I-III).

| $J^{PC}$ | I | II | III |
|---|---|---|---|
| $0^{++}$ | 45,94 | 41,69 | 42,09 |
| $1^{++}$ | 52,99 | 42,65 | 43,96 |
| $2^{++}$ | 58,46 | 48,04 | 49,34 |
| $0^{-+}$ | 45,89 | 35,51 | 36,98 |
| $1^{--}$ | 57,59 | 47,17 | 48,48 |
| $1^{+-}$ | 45,89 | 35,51 | 36,98 |

Table 3. Vertex functions

| $J^{PC}$ | $G_1^2$ |
|---|---|
| $0^{++}$ | $-8g/3$ |
| $1^{++}$ | $4g/3$ |
| $2^{++}$ | $4g/3$ |
| $0^{-+}$ | $8g/3 - 4g(m_i + m_k)^2/(3s_{ik})$ |
| $1^{--}$ | $4g/3$ |
| $1^{+-}$ | $8g/3 - 4g(m_i + m_k)^2/(3s_{ik})$ |

The vertex functions $G_1$ correspond to colour singlet states. $G_2^2(s_{ik}) = 3g$, here $g$ is the gluon constant. In the present paper the contribution of axial interaction to the states $J^{PC} = 0^{-+}$, $1^{+-}$ is taken into account.

Table 4. Coefficient of Chew-Mandelstam functions.

| $J^{PC}$ | $\alpha(J^{PC})$ | $\beta(J^{PC})$ | $\delta(J^{PC})$ |
|---|---|---|---|
| $0^{++}$ | $-1/2$ | $1/2$ | $0$ |
| $1^{++}$ | $1/2$ | $-e/2$ | $0$ |
| $2^{++}$ | $3/10$ | $1/5 - 3e/2$ | $-1/5$ |
| $0^{-+}$ | $1/2$ | $-e/2$ | $0$ |
| $1^{--}$ | $1/3$ | $1/6 - e/3$ | $-1/6$ |
| $1^{+-}$ | $1/2$ | $-e/2$ | $0$ |

Here is $e = (m_i - m_k)^2 / (m_i + m_k)^2$

Figure captions.

Fig. 1. Graphic representation of the equations for the four-quark subamplitudes $A_l(s, s_{12}, s_{123})$, $l = 1 - 4$. The coefficients are determined by the permutation of particles.

Fig. 2. Graphic representation of the equations for the glueball subamplitudes $A_p(s, s_{12}, s_{34})$, $p = 5 - 7$.

Fig. 3. Diagram of gluonic exchange defines the short-range component of quark interactions a) and box-diagram of meson M takes into account the long-range interaction component of the quark forces b).

Fig. 4. Contours of integration 1, 2 in the complex plane $s_{13}$ for the functions $J_1$, $J_3$.

Fig. 5. Contours of integration 1, 2, 3 in the complex plane $s_{134}$ (a) and $s_{13}$ (b) for the function $J_2$.

Fig. 6. Contours of integration 1, 2 in the complex plane $s_{24}$ for the function $J_3$.

## References.

1. Astier, A., et al: Phys. Lett. B25, 294 (1967)
2. Rosner, J.L.: Phys. Rev. Lett. 21, 950 (1968)
3. Harari, H.: Phys. Rev. Lett. 22, 562 (1969)
4. Jaffe, R.L.: Phys. Rev. D15, 267 (1977)
5. Chan Hong-Mo, Hogaasen, H.: Phys. Lett. B72, 121 (1977)
6. Wicklund, A.B. et al: Phys. Rev. Lett. 45, 1469 (1980)
7. Weinstein, J., Isgur, N.: Phys. Rev. Lett. 48, 659 (1982), Phys. Rev. D27 588 (1983)
8. Barnes, T.: Phys. Lett. B165, 434 (1985)
9. Godfrey, S., Isgur, N.: Phys. Rev. D32, 189 (1985)
10. Chodos, A., et al: Phys. Rev. D9, 3471 (1974)
11. Degrande, T., et al: Phys. Rev. D12, 2060 (1979)
12. Jaffe, R.L., Low, F.E.: Phys. Rev. D19, 2105 (1979)
13. De Rujula, A., Georgi, H., Glashow, S.L.: Phys. Rev. D12, 147 (1975)
14. Capstick, S., Isgur, N.: Phys. Rev. D34, 2809 (1986)
15. Gerasyuta, S.M.: Z. Phys. C60, 683 (1993)
16. Gerasyuta, S.M.: Nuovo Cim. A106, 37 (1993)
17. Gerasyuta, S.M., Kochkin, V.I.: Yad. Fiz. 59, 512 (1996)
18. Gerasyuta, S.M., Kochkin, V.I.: Z.Phys. C74, 325 (1997)
19. t'Hooft, G.: Nucl. Phys. B72, 461 (1974)
20. Veneziano, G.: Nucl. Phys. B117, 519 (1976)
21. Yakubovsky, O.A.: Yad. Fiz. 5, 1312 (1967)
22. Merkuriev, S.P., Faddeev, L.D.: Quantum scattering theory for systems of few particles. Moscow 1985.
23. Anisovich, V.V., Gerasyuta, S.M.: Yad. Fiz. 44, 174 (1986)
24. Anisovich, V.V., Gerasyuta, S.M., Sarantsev, A.V.: Int. J. Mod. Phys. A6, 625 (1991)
25. Chew, G.F., Mandelstam, S.: Phys. Rev. 119, 467 (1960)
26. Parisi, G., Petronzio, R.: Phys. Lett. B95, 51 (1980)
27. Konoplich, R., Shepkin, H.: Nuovo Cim. A67, 211 (1982)
28. Donoqhue, J.F.: Phys. Rev. D29, 2559 (1984)
29. Anisovich, V.V.: Proc. Int. Simp. Pion-Nucleon, Nucleon- Nucleon Physics. Gatchina. V.2, 237 (1989)
30. Gerasyuta, S.M., Keltuyala, I.V.: Yad. Fiz. 54, 793 (1991)
31. Particle Data Group. Phys. Rev. D54, 1 (1996)
32. Cornwall J.M., Soni A.: Phys. Lett. B120, 431 (1983)
33. Longacre R.S.: Phys. Rev. D42, 874 (1990)
34. Diekmann, B.: Phys. Rev. Lett. 159, 101 (1988)
35. Narison, S.: Z. Phys. C26, 209 (1984)
36. Carlson, C.E., et al: Phys. Rev. D30, 1594 (1984)
37. Achasov, N.N., Shestakov, G.N.: Usp. Phys. Nauk 161, 53 (1991)

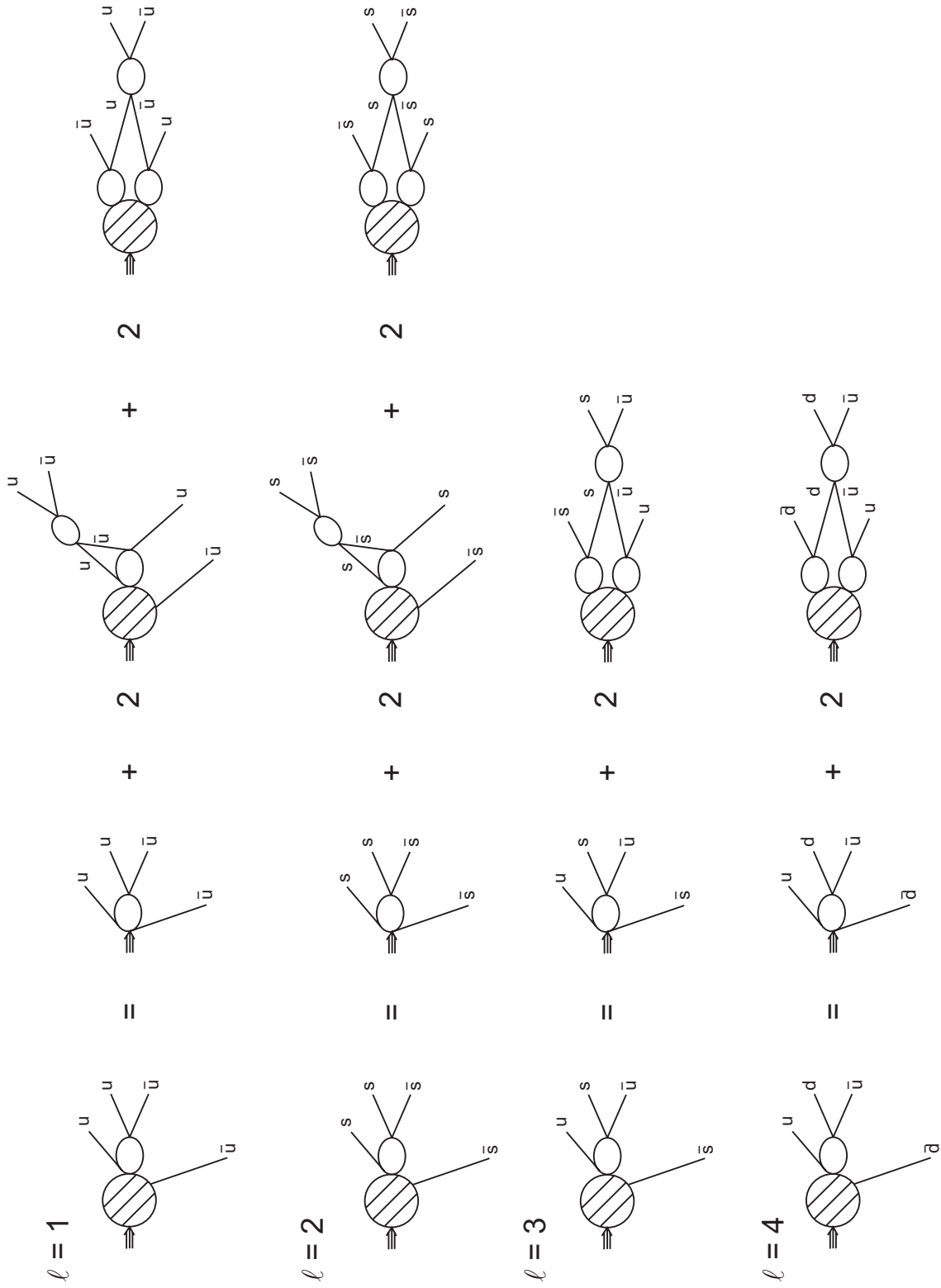

Fig. 1

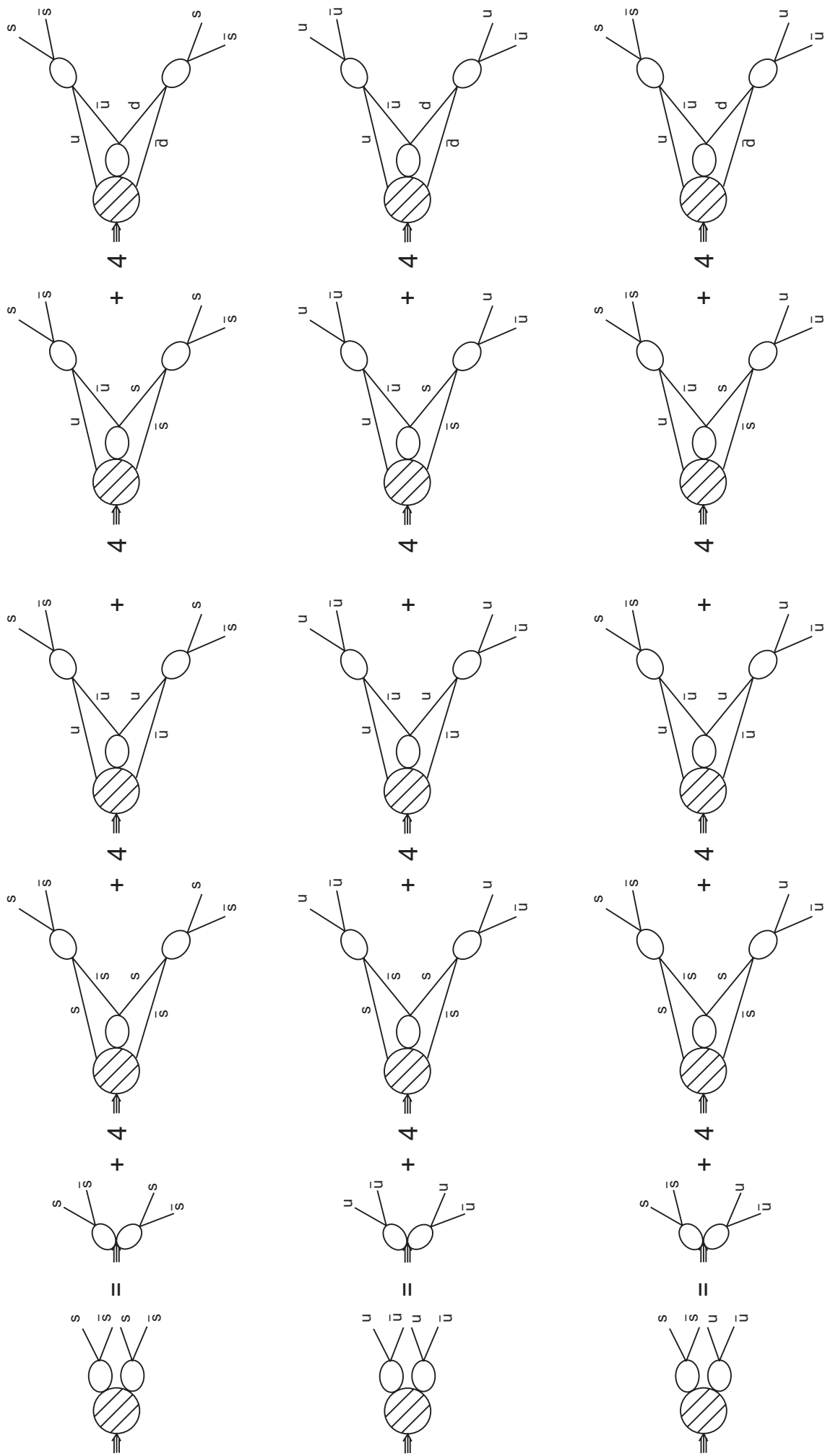

Fig. 2

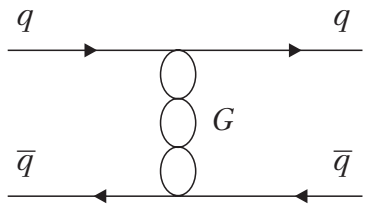

Fig. 3a

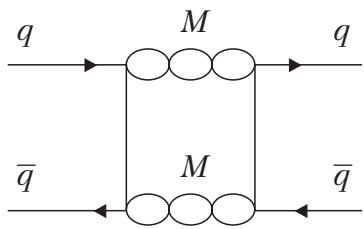

Fig. 3b

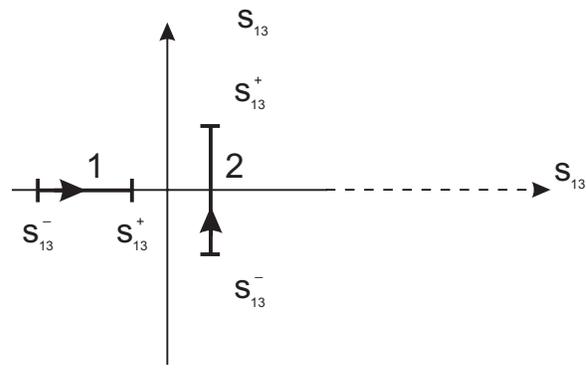

Fig. 4

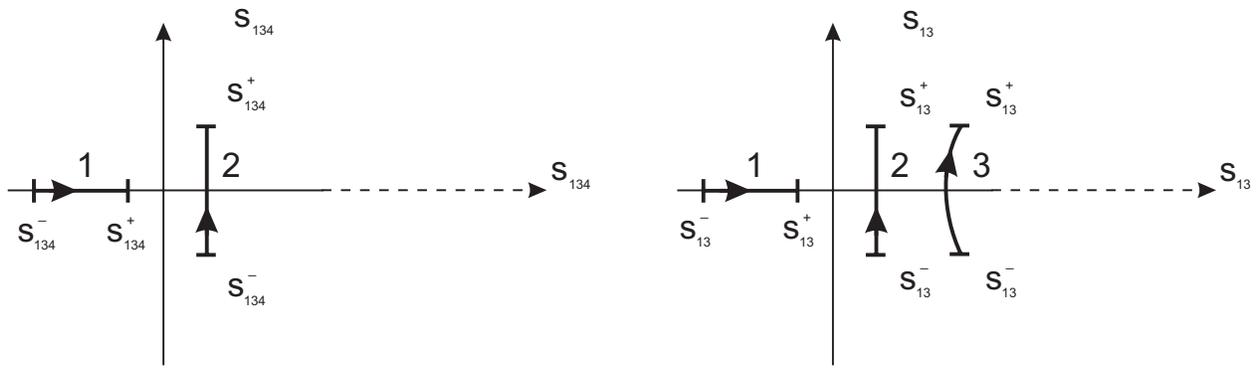

Fig. 5

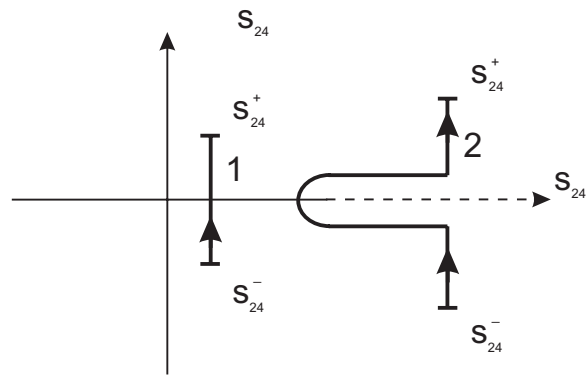

Fig. 6